\documentclass[conference]{IEEEtran}
\IEEEoverridecommandlockouts

\def\BibTeX{{\rm B\kern-.05em{\sc i\kern-.025em b}\kern-.08em
    T\kern-.1667em\lower.7ex\hbox{E}\kern-.125emX}}
  
\usepackage{graphicx}
\usepackage{cite}
\usepackage{amsmath,amssymb,amsfonts}
\usepackage{algorithmic}
\usepackage{textcomp}
\usepackage{xcolor}
\usepackage{booktabs}
\usepackage{svg}
\usepackage{xspace}
\usepackage[acronym]{glossaries}
\makeglossaries

\usepackage{booktabs} 
\usepackage{url}
\usepackage{color}
\usepackage{multirow}
\usepackage{enumitem}
\usepackage{array}
\usepackage[nolist,nohyperlinks]{acronym}
\usepackage{footnote}
\makesavenoteenv{tabular}
\usepackage{hyperref}
 
\newcommand{\resunetplusplus}{ResUNet++\xspace}

\acrodef{GI}{gastrointestinal}
\acrodef{CNN}{Convolutional Neural Network}
\acrodef{FCM}{Fuzzy C-mean Clustering}
\acrodef{ROI}{Region of Interest}
\acrodef{AI}{Artificial Intelligence} 
\acrodef{CAI}{Computer Assisted Intervention} 
\acrodef{ML}{Machine Learning}
\acrodef{IOU}[IoU]{Intersection over Union}
\acrodef{ROI}{Region of Interest}
\acrodef{CRC}{Colorectal Cancer}
\acrodef{CAD}{Computer-Aided Detection}
\acrodef{WCE}{Wireless Capsule Endoscopy}  
\acrodef{BoF}{Bag of Feature}  
\acrodef{GIANA}{Gastrointestinal Image ANAlysis} 
\acrodef{FCNN}{Fully Convolutional Neural Network}
\acrodef{FCN}{Fully Convolutional Network}
\acrodef{ASPP}{Atrous Spatial Pyramidal Pooling }
\acrodef{SGDR}{Stochastic Gradient Descent with Restart}
\acrodef{AUC-ROC}{Area Under Curve - Receiver Operating Characteristic}
\acrodef{ROC}{Receiver Operating Curve}
\acrodef{MSE}{Mean Square Error}
\acrodef{SGD}{Stochastic Gradient Descent}
\acrodef{BoF}{bag of feature}
\acrodef{NLP}{Natural Language Processing}
\acrodef{MSE}{Mean Square Error}
\acrodef{mIoU}{mean Intersection over Union}
\acrodef{ReLU}{Rectified Linear Unit}

\begin{document}

\title{\resunetplusplus: An Advanced Architecture for Medical Image Segmentation}

\author{\IEEEauthorblockN{Debesh Jha\IEEEauthorrefmark{1}\IEEEauthorrefmark{3}, Pia H. Smedsrud\IEEEauthorrefmark{1}\IEEEauthorrefmark{2}\IEEEauthorrefmark{4}, Michael A. Riegler\IEEEauthorrefmark{1}\IEEEauthorrefmark{4},  Dag Johansen\IEEEauthorrefmark{3}, \\ Thomas de Lange\IEEEauthorrefmark{2}\IEEEauthorrefmark{4}, P{\aa}l Halvorsen\IEEEauthorrefmark{1}\IEEEauthorrefmark{5},
H{\aa}vard D. Johansen\IEEEauthorrefmark{3}
}
\vspace{2mm}
\IEEEauthorblockA{\IEEEauthorrefmark{1}SimulaMet, Norway \ \ \ \ \ \
\IEEEauthorrefmark{2}Augere Medical AS, Norway \\
\IEEEauthorrefmark{3}UiT The Arctic University of Norway, Norway \ \ \ \ \ \
\IEEEauthorrefmark{4}University of Oslo, Norway\\
\IEEEauthorrefmark{5}Oslo Metropolitan University, Norway\\
Email: {debesh@simula.no}}}

\maketitle

\begin{abstract}
Accurate computer-aided polyp detection and segmentation during colonoscopy examinations can help endoscopists resect abnormal tissue and thereby decrease chances of polyps growing into cancer. Towards developing a fully automated model for pixel-wise polyp segmentation, we propose \resunetplusplus, which is an improved ResUNet architecture for colonoscopic image segmentation. Our experimental evaluations show that the suggested architecture produces good segmentation results on publicly available datasets. Furthermore, \resunetplusplus significantly outperforms U-Net and ResUNet, two key state-of-the-art deep learning architectures, by achieving high evaluation scores with a dice coefficient of 81.33\%, and a \ac{mIoU} of 79.27\% for the Kvasir-SEG dataset and a dice coefficient of 79.55\%, and a \ac{mIoU} of 79.62\% with CVC-612 dataset.

\end{abstract}


\vspace{-0.1cm}
\section{Introduction}
\label{sec:Introduction}
 
\ac{CRC} is one of the leading causes of cancer related deaths worldwide. Polyps are predecessors to this type of cancers and therefore important to discover early by clinicians through colonoscopy examinations.
To reduce the occurrence of \ac{CRC}, 
it is routine to resect the neoplastic lesions (for example, adenomatous polyps)~\cite{zauber2012colonoscopic}. 
Unfortunately, many adenomatous polyps are missed during the endoscopic examinations~\cite{van2006polyp}.
A \ac{CAD} system that, in real-time, can highlight the locations of polyps in the video stream from the endoscope, can act as a second observer, potentially drawing the endoscopist's attention to the polyps displayed on the monitor. This can reduce the chance that some polyps are overlooked~\cite{mori2018detecting}. For this purpose,  
an important improvement of pure anomaly detection approaches, which  only identify whether or not there is something abnormal in an image, we also want our \ac{CAD} system to have pixel-wise segmentation capability so that the specific regions of interest within each abnormal image can be identified.

\begin{figure}
 \centering
        \includegraphics[width=2cm, height=2cm]{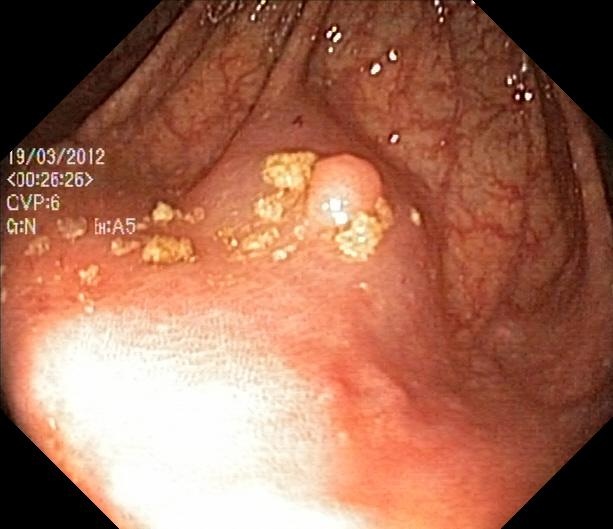}
        \includegraphics[width=2cm, height=2cm]{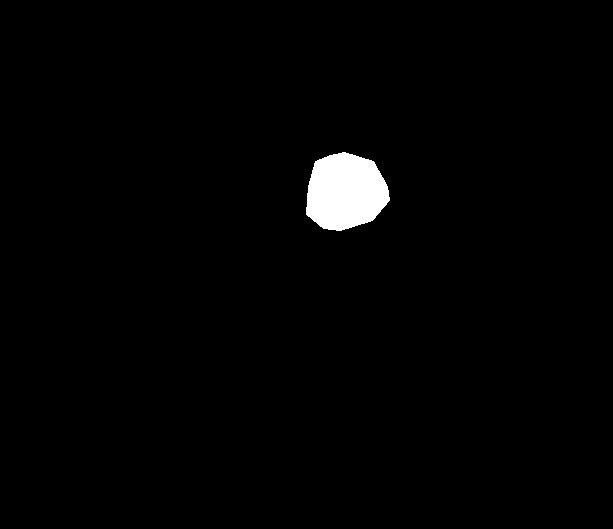}
        \includegraphics[width=2cm, height=2cm]{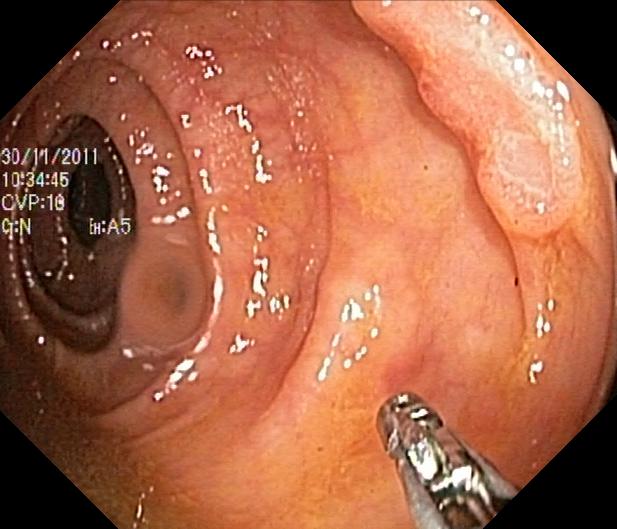}
        \includegraphics[width=2cm, height=2cm]{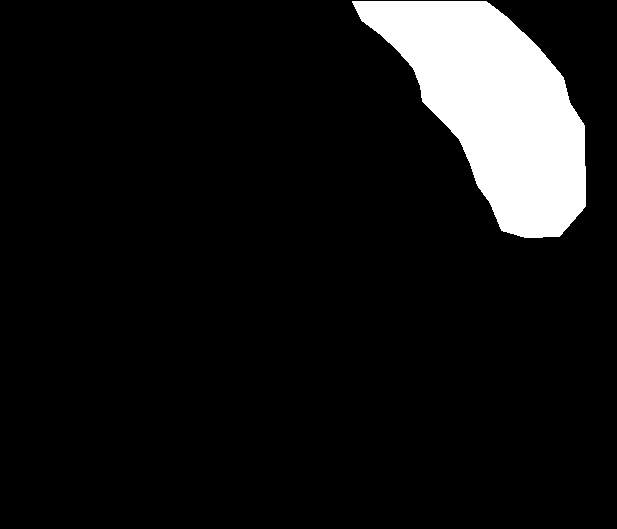} \\
        \vspace{1mm}
        \includegraphics[width=2cm, height=2cm]{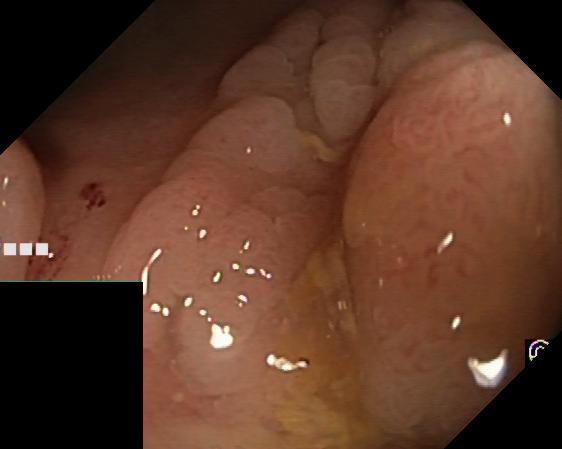}
        \includegraphics[width=2cm, height=2cm]{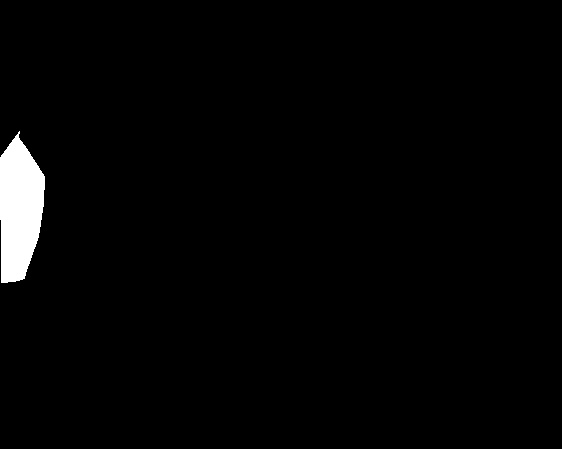}
        \includegraphics[width=2cm, height=2cm]{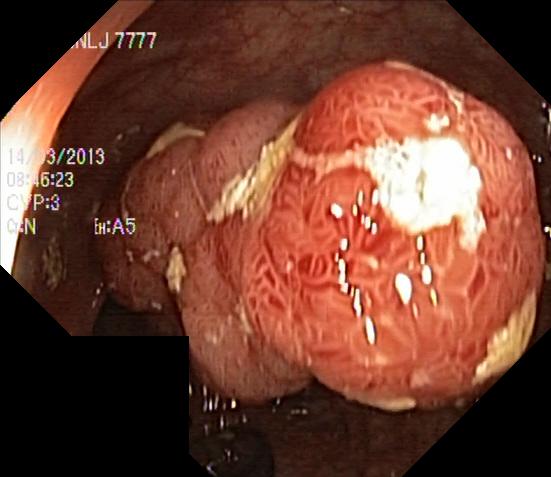}
        \includegraphics[width=2cm, height=2cm]{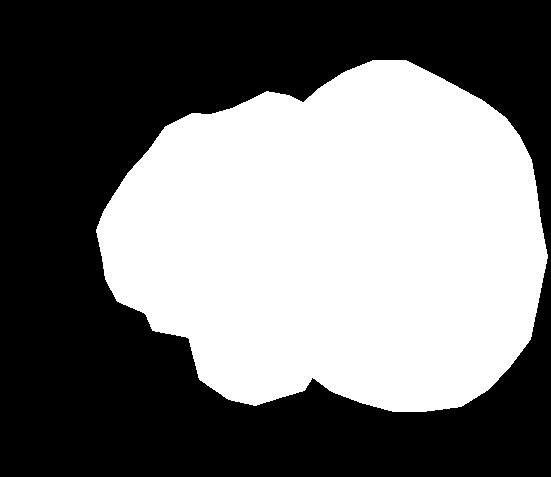}
    
    \caption{Examples of polyp images and their corresponding masks from Kvasir-SEG dataset. The first and third column represents the original images, and the second column and fourth column represents their corresponding ground truth.}
    \label{fig:polyp_with_mask}
      \vspace{-5mm}
\end{figure}

A key challenge for designing a precise \ac{CAD} system
for polyps is the high costs of collecting and labeling proper medical datasets for training and testing. 
Polyps  come in a wide variety of shapes, sizes, colors, and appearances as shown  in Figure~\ref{fig:polyp_with_mask}. 
For the four main classes of polyps: adenoma, serrated, hyperplastic, and mixed (rare), there are high inter-class similarity and intra-class variation. 
There can also be high background object similarity, for instance, where parts of a polyp is covered with stool or when they blend into the background mucosa. Although these factors make our task challenging, we conjecture that 
there is still a high potential for designing a system with a performance acceptable for clinical use. 

Motivated by the recent success of semantic segmentation-based approaches for medical image analysis~\cite{milletari2016v,ronneberger2015u,zhang2018road}, we explore how these methods can be used to improve the performance for automatic polyp segmentation and detection. A popular deep learning architecture in the field of semantic segmentation for biomedical application is U-Net~\cite{ronneberger2015u}, which have shown state-of-the-art performance at the $2015$ ISBI cell tracking challenge~\footnote{ \href{http://brainiac2.mit.edu/isbi_challenge/}{http://brainiac2.mit.edu/isbi\_challenge/}.}. The ResUNet~\cite{zhang2018road} architecture, is a variant of U-Net architecture that has provided state-of-the-art results for the road image extraction. We therefore adapt this architecture as a basis for our work.

In this paper, we propose the \resunetplusplus architecture for medical image segmentation. We have evaluated our model on two publicly available datasets. Our experimental results reveal that the improved model is efficient and achieved a performance boost compared to the popular U-Net~\cite{ronneberger2015u} and ResUNet~\cite{zhang2018road} architectures.

In summary, the contributions of the paper are as follows:
\begin{enumerate}
\item We propose the novel \resunetplusplus architecture, which is a semantic segmentation neural network that takes advantage of residual blocks, squeeze and excitation blocks, \ac{ASPP}, and attention blocks. \resunetplusplus improved the segmentation results significantly for the colorectal polyps compared to other state-of-the-art methods. The proposed architecture works well with a smaller number of images.  

\item We annotated the polyp class from the Kvasir dataset~\cite{pogorelov2017kvasir} with the help of an expert gastroenterologist to create the new Kvasir-SEG dataset~\cite{debeshkvasir-SEG}. 
We make this polyp segmentation dataset available to the research community to foster development of new methods and  reproducible research. 

\end{enumerate}
\section{Related Work}
\label{section:related_work}

Automatic~\ac{GI} tract disease detection and classification in colonoscopic videos has been an active area of research for the past two decades.
Polyp detection has in particular been given attention.
The performance of the machine learning software has come close to the level of expert endoscopists~\cite{wang2014part,mori2017computer,brandao2018towards,wang2018development}.

Apart from work on algorithm development, researchers have also investigated complete \ac{CAD} systems, from data annotation, analysis, and evaluation to visualization for the medical experts~\cite{wang2015polyp, riegler2017annotation,hicks2018mimir}. Thambawita et al.~\cite{thambawita2018medico} explored various methods, ranging from \ac{ML} to deep \ac{CNN}, and suggested five novel models as a potential solution for classifying \ac{GI} tract findings into sixteen classes. Guo et al.~\cite{guo2019giana} presented two variants of fully convolutional neural networks, which secured the first position at the 2017~\ac{GIANA} challenge and second position at the 2018~\ac{GIANA} challenge. 

Long et al.~\cite{long2015fully} proposed a state-of-the-art semantic segmentation approach for image segmentation known as a~\ac{FCN}. \ac{FCN} are trained end-to-end, pixels-to-pixels, and outputs segmentation result without any additional post-processing steps. Ronneberger et al.~\cite{ronneberger2015u} modified and extended the FCN architecture to an U-Net architecture. There are various modification and extension based on~\mbox{U-Net} architecture~\cite{cciccek20163d,milletari2016v,drozdzal2016importance,zhang2018road,diakogiannis2019resunet,guo2019giana,zhou2018unet++} to achieve better segmentation results on both natural images and biomedical images.

Most of the published work in the field of polyp detection perform well on the specific datasets, and test scenarios often used small training and validation datasets~\cite{wang2013part,brandao2018towards}. The model evaluated on the smaller dataset is neither generalizable nor robust. Moreover, some of the research work only focus on a specific type of polyps. Some of the current work also use non-publicly datasets, which makes it difficult to compare and reproduce results. Therefore, the goal of the \ac{ML} models to reach a performance level similar to, or better than colonoscopists has not been achieved yet. There exists a potential for improvement in boosting the performance of the system.

 \begin{figure}[!ht]
   \centering
  \includegraphics[trim=1.3cm 2.5cm 2.5cm 1.1cm, clip, width=\linewidth]{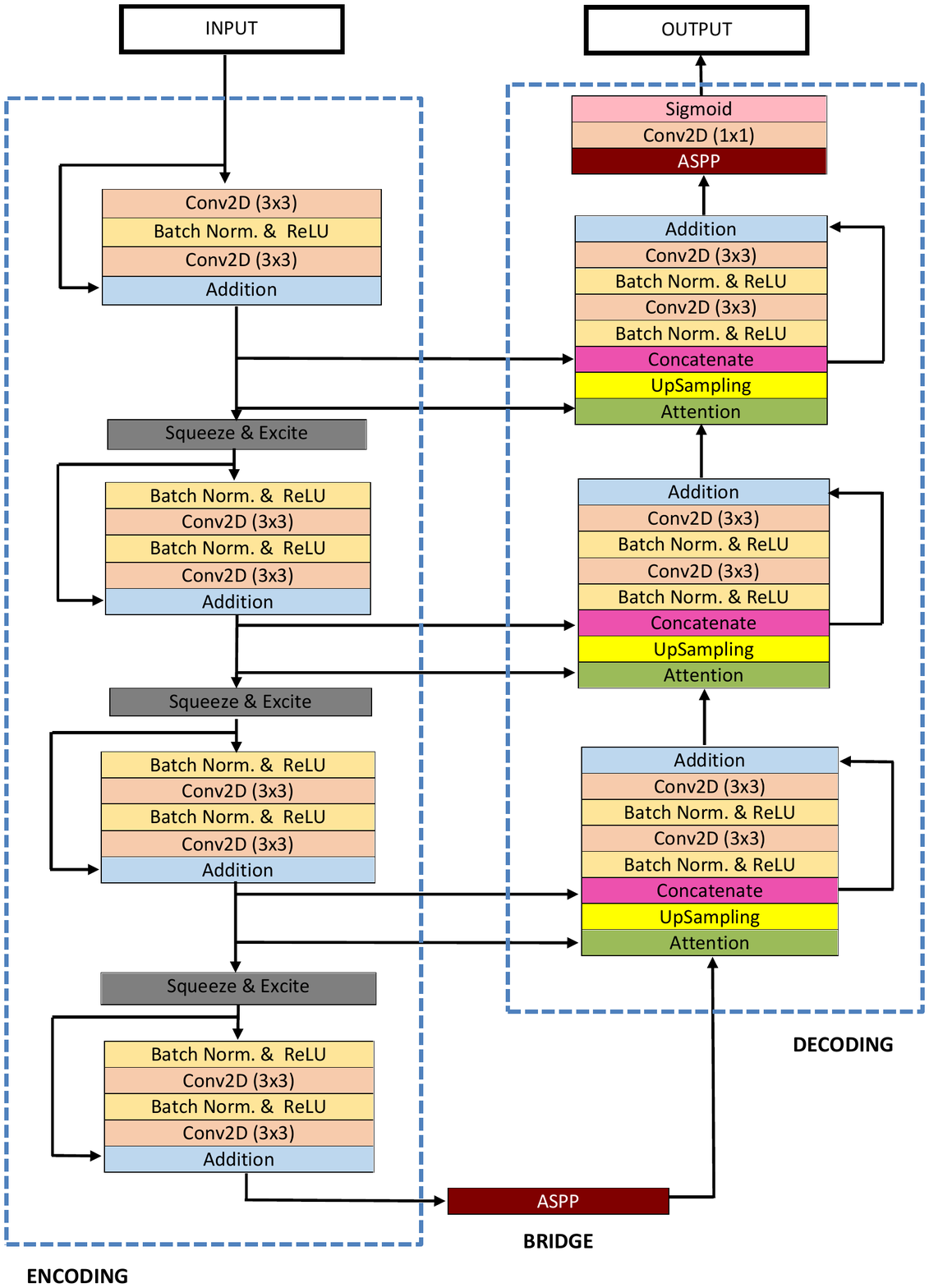}
 \caption{Block diagram of the proposed \resunetplusplus architecture.}
  \label{fig:proposed_method}
  \vspace{-5mm}
 \end{figure}

\section{\resunetplusplus}
\label{section:resunetplusplus}

The \resunetplusplus architecture is based on the Deep Residual U-Net (ResUNet)~\cite{zhang2018road}, which is an architecture that uses the strength of deep residual learning~\cite{he2016deep} and U-Net~\cite{ronneberger2015u}. The proposed \resunetplusplus architecture takes advantage of the residual blocks, the squeeze and excitation block, \ac{ASPP}, and the attention block. 

The residual block propagates information over layers, allowing to build a deeper neural network that could solve the degradation problem in each of the encoders. This improves the channel inter-dependencies, while at the same time reducing the computational cost. The proposed \resunetplusplus architecture contains one stem block followed by three encoder blocks, \ac{ASPP}, and three decoder blocks.  The block diagram of the proposed \resunetplusplus architecture is shown in Figure~\ref{fig:proposed_method}. In the block diagram, we can see that the residual unit is a combination of batch normalization, \ac{ReLU} activation, and convolutional layers. 

Each encoder block consists of two successive $3 \times 3$ convolutional block and an identity mapping. Each convolution block includes a batch normalization layer, a \ac{ReLU} activation layer, and a convolutional layer. The identity mapping connects the input and output of the encoder block. A strided convolution layer is applied to reduce the spatial dimension of the feature maps by half at the first convolutional layer of the encoder block. The output of encoder block is passed through the squeeze-and-excitation block. The \ac{ASPP} acts as a bridge, enlarging the field-of-view of the filters to include a broader context. Correspondingly, the decoding path consists of residual units, too. Before each unit, the attention block increases the effectiveness of feature maps. This is followed by a nearest-neighbor up-sampling of feature maps from the lower level and the concatenation with feature maps from their corresponding encoding path.

The output of the decoder block is passed through \ac{ASPP}, and finally, we apply a $1 \times 1$ convolution with sigmoid activation, that provides the segmentation map. The extension of the \resunetplusplus is the squeeze-and-excitation blocks marked in light blue, the \ac{ASPP} block marked in dark red, and attention block marked in light green. A brief explanation of each of the parts is given in the following subsections.

 \subsection{Residual Units}
Deeper Neural Networks are comparatively challenging to train. Training a deep neural network with an increasing network depth can improve accuracy. However, it can hamper the training process and cause a degradation problem~\cite{he2016deep,zhang2018road}. He et al.~\cite{he2016deep} proposed a deep residual learning framework to facilitate the training process and address the problem of degradation. ResUNet~\cite{zhang2018road} uses full pre-activation residual units. The deep residual unit makes the deep network easy to train and the skip connection within the networks helps to propagate information without degradation, improving the design of the neural network by decreasing the parameters along with comparable performance or boost in performance on semantic segmentation task~\cite{he2016deep,zhang2018road}. Because of these advantages, we use ResUNet as the backbone architecture.  

 \subsection{Squeeze and Excitation Units}
 The squeeze-and-excitation network~\cite{hu2018squeeze} boosts the representative power of the network by re-calibrating the features responses employing precise modeling inter-dependencies between the channels. The goal of the squeeze and excite block is to ensure that the network can increase its sensitivity to the relevant features and suppress the unnecessary features. This goal is achieved in two steps. The first step is squeeze (global information embedding), where each channel is squeezed by using global average pooling for generating channel-wise statistics. The second step is excitation (active calibration) that aims to capture the channel-wise dependencies fully~\cite{hu2018squeeze}. In the proposed architecture, the squeeze and excitation block is stacked together with the residual block to increase effective generalization over different datasets and improve the performance of the network.  

 \subsection{Atrous Spatial Pyramidal Pooling}
The idea of~\ac{ASPP} comes from spatial pyramidal pooling~\cite{he2015spatial}, which is successful at re-sampling features at multiple scales. In~\ac{ASPP}, the contextual information is captured at various scales~\cite{chen2018deeplab,chen2017rethinking} and many parallel atrous convolutions~\cite{chen2017deeplab} with different rates in the input feature map are fused. Atrous convolution allows controlling the field-of-view for capturing multi-scale information precisely. In the proposed architecture, \ac{ASPP} acts as a bridge between encoder and decoder in our architecture, as shown in Figure~\ref{fig:proposed_method}. The~\ac{ASPP} model has shown promising results on various segmentation tasks by providing multi-scale information. Therefore, we use~\ac{ASPP} to capture the useful multi-scale information for the semantic segmentation task. 

\subsection{Attention Units}
The attention mechanism is mostly popular in~\ac{NLP}~\cite{vaswani2017attention}. It gives attention to the subset of its input. Moreover, it has been employed in semantic segmentation tasks, like pixel-wise prediction~\cite{li2018pyramid}. The attention mechanism determines which parts of the network require more attention in the neural network. The attention mechanism also 
reduces the computational cost of encoding 
the information in each polyp image into a vector of fixed dimension. The main advantage of the attention mechanism is that they are simple, can be applied to any input size, enhance the quality of features that boosts the results. 

In the previous two approaches, U-Net~\cite{ronneberger2015u} and ResUNet~\cite{zhang2018road}, there exists a direct concatenation of the encoder feature maps with the decoder feature maps. Inspired by the success of attention mechanism, both in~\ac{NLP} and computer vision tasks, we implemented the attention block in the decoder part of our architecture to be able to focus on the essential areas of the feature maps. 
\section{Experiments}
\label{section:Experiment}

To evaluate the \resunetplusplus architecture, we train, validate, and test models using two publicly available datasets. We compare the performance of our \resunetplusplus models with ones trained using U-Net and ResUNet.
 
\subsection{Datasets}
For the task of polyp image segmentation, each pixel in the training images must be labeled as belonging to either the polyp class or the non-polyp class. For the evaluation of \resunetplusplus, we use the Kvasir-SEG dataset~\cite{debeshkvasir-SEG}, which consists of 1,000 polyp images and their corresponding ground truth masks annotated by expert endoscopists from Oslo University Hospital (Norway). Example images and their corresponding masks from the Kvasir-SEG dataset are shown in Figure~\ref{fig:polyp_with_mask}. The second dataset we have used is the CVC-ClinicDB database~\cite{bernal2015wm}, which is an open-access dataset of $612$ images with a resolution of $384 \times 288$ from 31 colonoscopy sequences.
\subsection{Implementation details}
All architectures were implemented using the Keras framework~\cite{chollet2015keras} with TensorFlow~\cite{abadi2016tensorflow} as backend. We performed our experiment on a single Volta 100 GPU on a powerful Nvidia DGX-2 AI system capable of 2-petaFLOPS tensor performance. The system is part of Simula Research Laboratories heterogeneous cluster and has dual Intel(R) Xeon(R) Platinum 8168 CPU@2.70GHz, 1.5TB of DDR4-2667MHz DRAM, 32TB of NVMe scratch space, and 16 of NVIDIAs latest Volta 100 GPGPUs interconnected using Nvidia's NVlink fully non-blocking crossbars switch capable of 2.4 TB/s of bisectional bandwidth. The system was running Ubuntu 18.04.3LTS OS and had the latest Cuda 10.1.243 installed. We start the training with a batch size of 16, and the proposed architecture is optimized by Adam optimizer. The learning rate of the algorithm is set to $1\mathrm{e}{-4}$. A lower learning rate is preferred, although a lower learning rate slowed down convergence, and a larger learning rate often causes convergence failures. 

The size of the image within the same dataset varies. Both the dataset used in the study consists of different resolution images. For efficient GPU utilization and to reduce the training time, we crop the images by putting a crop margin of $320 \times 320$ to increase the training dataset. Then, the images are resized to $256 \times 256$ pixels before feeding the images to the model. We have used the data augmentation technique such as center crop, random crop, horizontal flip, vertical flip, scale augmentation, random rotation, cutout, and brightness augmentation, etc., to increase the number of training samples. The rotation angle is randomly chosen from $0$ to $90$\textdegree. We have utilized $80$\% of the dataset for training, $10$\% for validation, and $10$\% for the testing. We trained all the models for $120$ epochs with a lower learning rate so that a more generalized model can be built. The batch size, epoch, and learning rate were reset depending upon the need. There was an accuracy trade-off if we decrease the batch size; however, we preferred a larger batch size over accuracy because smaller batch size can lead to over-fitting. We also used the \ac{SGDR} to improve the performance of the model.

\section{Results}
\label{section:Results}
To show the effectiveness of \resunetplusplus, we conducted two sets of experiments on Kvasir-SEG and CVC-612 datasets. For the model comparison, we compared the results of the proposed \resunetplusplus with the original U-Net and original ResUNet architecture, as both of them are the common preference for the semantic segmentation task. The original implementation of ResUNet, which uses \ac{MSE} as the loss function, did not produce satisfactory results with Kvasir-SEG and CVC-612 datasets. Therefore, we replaced the \ac{MSE} loss function with dice coefficient loss and did hyperparameter optimization to improve the results and named the architecture as ResUNet-mod. With this modification, we achieved a performance boost in ResUNet-mod architecture for both the datasets.

\subsection{Results on the Kvasir-SEG dataset}
We have tried different sets of hyperparameters (i.e., learning rate, number of epochs, optimizer, batch size, and filter size) for the optimization of \resunetplusplus architecture. Hyperparameter tuning is done manually by training the models with different sets of hyperparameters and evaluating their results. The results of \resunetplusplus, ResUNet-mod, ResUNet~\cite{zhang2018road}, and U-Net~\cite{ronneberger2015u} are  presented in Table~\ref{table:results}.  Table~\ref{table:results} shows that the proposed model achieved the highest dice coefficient, \ac{mIoU}, recall, and competitive precision for the Kvasir-SEG dataset. U-Net achieved the highest precision. However, the dice coefficient and \ac{mIoU} scores are not competitive, which is an important metric for semantic segmentation task. The proposed architecture has outperformed the baseline architectures by a significant margin in terms of \ac{mIoU}. 

\begin{table}[t]
 \caption{The table shows the evaluation results of all the models on Kvasir-SEG dataset.}
    \label{table:results}
    \vspace{-5mm}
   \def\arraystretch{1.1}
    \setlength\tabcolsep{5pt}
    \par\bigskip
    \centering
        \resizebox{\columnwidth}{!}{%
           \begin{tabular}{ l c c c c } 
                \bottomrule
                Method & Dice & mIoU & Recall & Precision\\ 
                 \bottomrule
                \textbf{ResUNet++} & \textbf{{0.8133}}  & \textbf{{0.7927} }& \textbf{0.7064} &  0.8774\\ 
                {ResUNet-mod} & 0.7909 &  0.4287 &  0.6909  &  0.8713\\ 
                {ResUNet} & 0.5144 &  0.4364 &  0.5041  &  0.7292\\ 
                {U-Net} & 0.7147 & 0.4334 &   0.6306 & \textbf{0.9222 }\\  
                  \toprule
                    \vspace{-5 mm}
\end{tabular}}
\end{table}									
\begin{table}[b]
 \caption{The table shows the evaluation results of all the models on CVC-612 dataset.}
    \label{table:results1}
    \vspace{-5mm}
   \def\arraystretch{1.1}
    \setlength\tabcolsep{5pt}
    \par\bigskip
    \centering
        \resizebox{\columnwidth}{!}{%
           \begin{tabular}{ l c c c c } 
                \bottomrule
                Method & Dice & mIoU & Recall & Precision\\ 
               \bottomrule
                \textbf{ResUNet++} & \textbf{0.7955} &  \textbf{0.7962} & \textbf{0.7022} &  0.8785\\ 
                {ResUNet-mod} & 0.7788  & 0.4545  & 0.6683 & \textbf{0.8877}\\ 
                {ResUNet} & 0.4510  & 0.4570  & 0.5775 & 0.5614\\ 
                {U-Net} & 0.6419 & 0.4711 & 0.6756   & 0.6868\\ 
                  \toprule
                    \vspace{-10pt}
\end{tabular}}
\end{table}									

\begin{figure*}[!t]
    \centering
    \includegraphics [height=1.8cm]{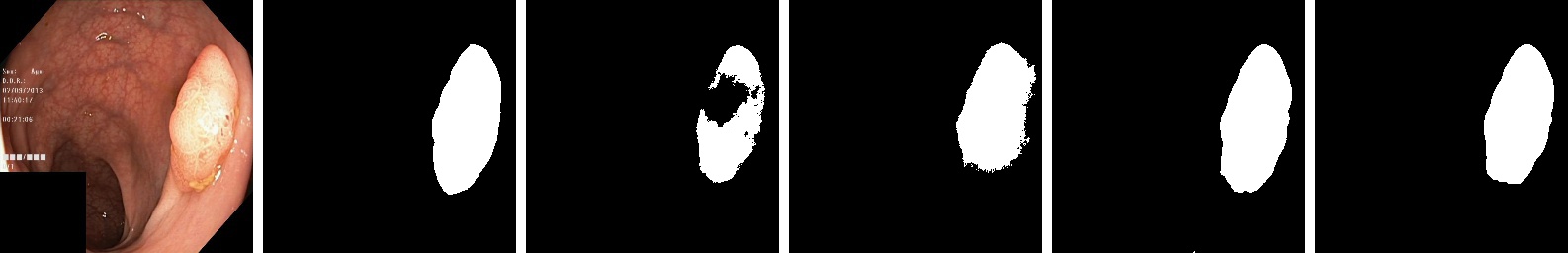}\vspace{1mm}\
    \includegraphics [height=1.8cm]{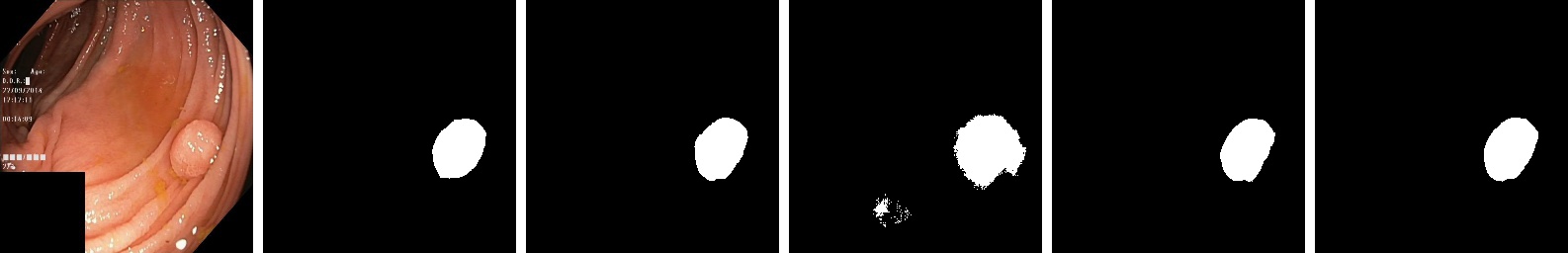}\vspace{1mm}\
    \includegraphics [height=1.8cm]{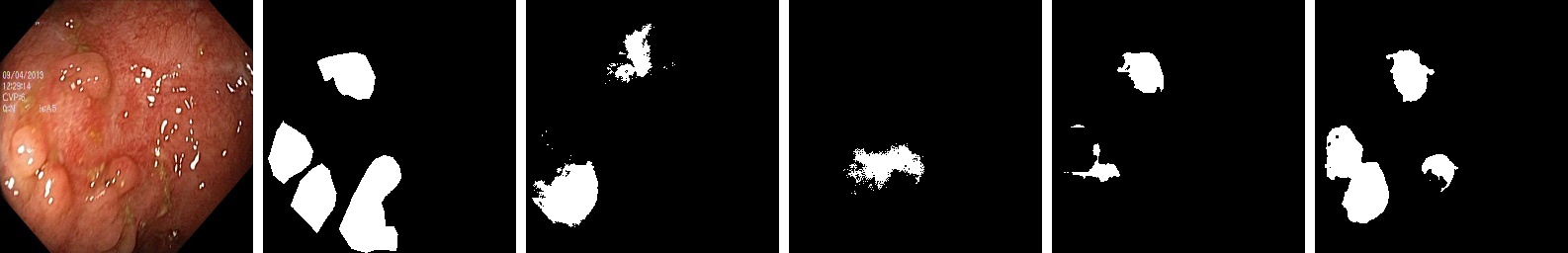}\vspace{1mm}\
    \includegraphics [height=1.8cm]{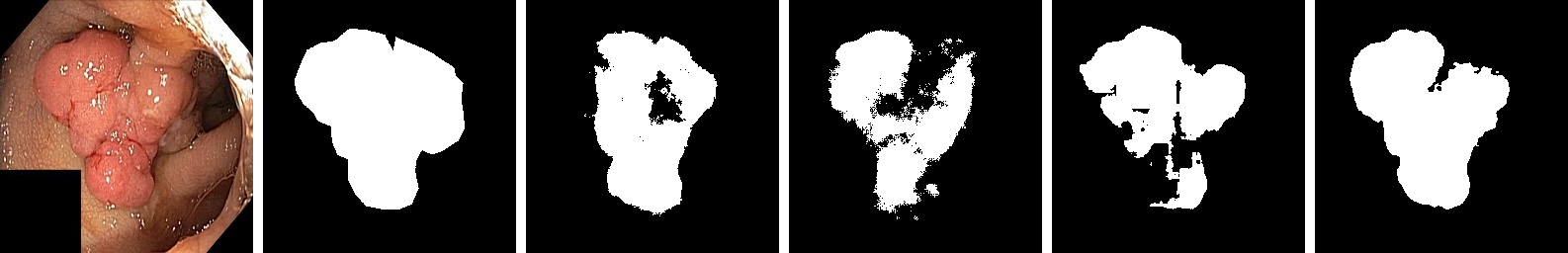}\vspace{1mm}\
    \includegraphics [height=1.8cm]{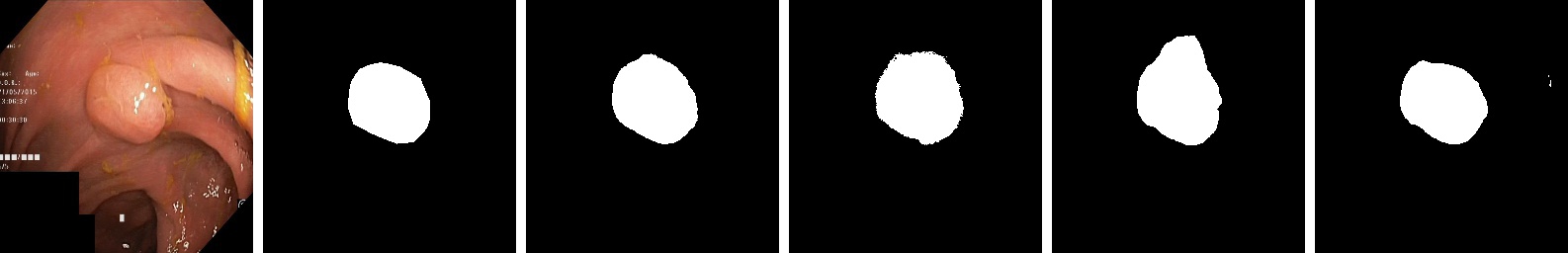}\vspace{1mm}\
    \includegraphics [height=1.8cm]{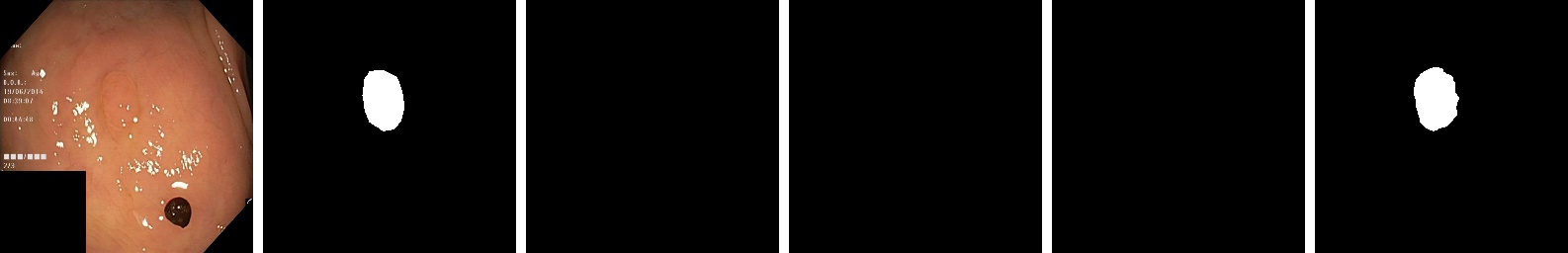}\vspace{1mm}
    \caption{Qualitative results comparison on the Kvasir-SEG dataset. From the left: image (1), (2) Ground truth, (3) U-Net, (4) ResUNet, (5) ResUNet-mod, and (6) \resunetplusplus. From the experimental results, we can say that \resunetplusplus produces better segmentation masks than other competitors.}
  \label{fig:comparison_with_baseline}
      \vspace{-5mm}
\end{figure*}

\subsection{Results on the CVC-612 dataset}
We have performed additional experiments for in-depth performance analysis for automatic polyp segmentation. Therefore, we attempted for the generalization of the model to check the generalizability capability of the proposed architecture on a different dataset. Generalizability would be a further step toward building clinical acceptable model. Table~\ref{table:results1} shows the results for all the architectures on CVC-612 datasets. The proposed model obtained highest dice coefficient, \ac{mIoU}, and recall and competitive precision. \\ 

Figure~\ref{fig:comparison_with_baseline} shows the qualitative results for all the models. From Table~\ref{table:results}, Table~\ref{table:results1}, and Figure~\ref{fig:comparison_with_baseline} we demonstrate the superiority of \resunetplusplus over the baseline architectures. The quantitative and qualitative result shows that the \resunetplusplus model trained on Kvasir-SEG and CVC-612 dataset performs well and outperforms all other models in terms of dice coefficient, \ac{mIoU}, and recall. Therefore, the \resunetplusplus architecture should be considered over these baselines architecture in the medical image segmentation task.


\section{Discussion}
\label{discussion}

The proposed \resunetplusplus architecture produces satisfactory results on both Kvasir-SEG and CVC-612 datasets. From Figure~\ref{fig:comparison_with_baseline}, it is evident that the segmentation map produced by \resunetplusplus outperforms other architectures in capturing shape information, in the Kvasir-SEG dataset. It means the generated segmentation mask in \resunetplusplus has more similar ground truth then the presented state-of-the-art models. However, ResUNet-mod and U-Net also produced competitive segmentation masks.

We trained the model using different available loss functions, for example, binary cross-entropy, the combination of binary cross-entropy and dice loss, and mean square loss. We observed that the model achieved a higher dice coefficient value with all the loss function. However, \ac{mIoU} were significantly lower with all other except dice coefficient loss function. We selected the dice coefficient loss function based on our empirical evaluation. Moreover, we also observed that the number of filters, batch size, optimizer, and loss function can influence the result.  

We conjecture that the performance of the model can be further improved by increasing the dataset size, applying more augmentation techniques, and by applying some post-processing steps. Despite increased numbers of parameters with the proposed architecture, we trained the model to achieve higher performance. We conclude that the application of \resunetplusplus should not only limited to biomedical image segmentation but could also be expanded to the natural image segmentation and other pixel-wise classification tasks, which need further detailed validations. We have optimized the code as much as possible based on our knowledge and experience. However, there may exist further optimization, which may also influence the results of the architectures. We have run the code only on a Nvidia-DGX-2 machine, and the images were resized, which may have lead to the loss of some useful information. Additionally, \resunetplusplus uses more parameters, which increases training time.

\section{Conclusion}
\label{section:conclusion}
In this paper, we presented \resunetplusplus, which is an architecture to address the need for more accurate segmentation of colorectal polyps found in colonoscopy examinations. The suggested architecture takes advantage of residual units, squeeze and excitation units, \ac{ASPP}, and attention units. Comprehensive evaluation using different available datasets demonstrates that the proposed \resunetplusplus architecture outperforms the state-of-the-art U-Net and ResUNet architectures in terms of producing semantically accurate predictions. Towards achieving the generalizability goal, the proposed architecture can be a strong baseline for further investigation in the direction of developing a clinically useful method. Post-processing techniques can potentially be applied to our model to achieve even better segmentation results.
\section*{Acknowledgement}
This work is funded in part by Research Council of Norway project number 263248. The computations in this paper were performed on equipment provided by the Experimental Infrastructure for Exploration of Exascale Computing (eX3), which is financially supported by the Research Council of Norway under contract 270053.

\bibliographystyle{IEEEtran}
\bibliography{references} 
\end{document}